# Temperature-sensitive spin-wave nonreciprocity induced by interlayer dipolar coupling in ferromagnet/X (X=paramagnet, superconductor) hybrid systems

M.A. Kuznetsov,[1,*] A.A. Fraerman[1,†]

[1]*Institute for Physics of Microstructures, Russian Academy of Sciences, Akademicheskaya St. 7,*

*Nizhny Novgorod 607680, Russian Federation*

Spin-wave (SW) spectra have theoretically been studied in a thin film of a ferromagnet (FM) on a substrate from a paramagnet (PM) (an FM above the critical temperature) or from superconductor (SC). A spin wave propagating in the FM induces the dynamic magnetization and superconducting current in the underlying PM and SC, respectively, which affect the SW propagation by their magnetic fields. As a result of this interaction, the SW spectrum becomes nonreciprocal to depend on the sign of the SW wave vector $\mathbf{q}$. We show that the nonreciprocal contribution to the SW spectra in FM/PM and FM/SC systems is given by the frequency shift of $\Delta\omega(\mathbf{q}) \equiv \omega(\mathbf{q}) - \omega(-\mathbf{q}) = a(T)(\boldsymbol{\tau} \cdot \mathbf{q})$ with $\boldsymbol{\tau} = (\mathbf{n} \times \mathbf{M})$ being the toroidal magnetic moment, $\mathbf{M}$ the FM magnetization, $\mathbf{n}$ the unit vector normal to the FM/PM(SC) interface, and $a(T)$ the temperature-dependent constant of a dipole nature, whose sign depends on the substrate type. As the $\Delta\omega(T)$ dependence is strong at temperatures $T$ close to the critical temperature $T_c$ for the FM-PM or normal metal-SC transition, one gets a possibility to control the frequency SW nonreciprocity with temperature variation near $T_c$. The dipolar mechanism we propose for SW frequency nonreciprocity is promising for introducing this property of SW propagation into functional devices.

## I. INTRODUCTION

Spin-wave (SW) nonreciprocity, that is, dependence of the SW properties on the sign of SW wave vector, $\mathbf{q}$, is an interesting feature of SW propagation in ferromagnets (FM´s) [1]. Nonreciprocal devices of this kind are being widely used in microwave technology [2–7]. The well-known paradigm for SW nonreciprocity is the surface magnetostatic Damon-Eshbach mode occurring in an FM film, whose SW amplitude depends on propagation direction [8–11]. On the other hand, SW nonreciprocity may manifest in the frequency shift of spin waves with the same wavelength. Phenomenologically, the frequency SW nonreciprocity is described by a term that breaks the time-reversal symmetry of the spectrum,

$$\omega(\mathbf{q}) = \cdots + \frac{a}{2}(\boldsymbol{\tau} \cdot \mathbf{q}) + \cdots, \qquad (1)$$

where $a$ is the phenomenological constant, which can generally depend on the $q$, $\boldsymbol{\tau} = (\mathbf{n} \times \mathbf{M})$ is the toroidal magnetic moment [12], $\mathbf{M}$ is the magnetization vector, and $\mathbf{n}$ is a unit vector normal to the film plane. Note that the term in Eq. (1) can be constructed uniquely from the vectors **n**, **M** and **q**, and its sign depends on the sign of **q**. This term also exists only in films with dissimilar boundaries; otherwise, the vector **n** cannot be selected. The dissimilarity of the film boundaries can be achieved, for instance, by covering one of them by a metal layer with thickness comparable to its skin depth [9–11]. In such a system, the frequency SW nonreciprocity results from the interaction of a spin wave with the conduction current induced in the metal cover by it.

---

[*]kuznetsovM@ipmras.ru

[†]andr@ipmras.ru

In our work we consider another origin of the SW frequency nonreciprocity. It stems from interlayer dipolar coupling in a hybrid system that consists of a thin FM film on a substrate from a paramagnet (PM) [13,14] or from a superconductor (SC) [14–16]. The FM/PM(SC) system breaks out the chiral symmetry, that is, in the FM/PM [13] and in the FM/SC [16] systems, the energies of the left and the right cycloidal magnetic spirals are different. So, in the FM/PM and the FM/SC, the right and the left spirals have lowest energy, respectively [13,16]. The case of spin waves is similar to these static magnetic distributions, that is the snapshots of the counter propagating spin waves are the right [Fig. 1(a),(d)] and the left [Fig. 1(b),(c)] magnetic spirals mentioned above. Thus the directions for preferable SW propagation, corresponding to the minimum of the dipolar energy, are opposite for PM and SC (see panel 1 in Fig. 1), which reflects the dependence of the sign of the constant $a$ in Eq. (1) on the type of substrate, PM or SC. This feature can be understood, in particular, by considering the system with an ideal PM (infinitely high susceptibility) or an ideal SC (the Meissner state), for which the tangential (for PM) or normal (for SC) component of a magnetic field should vanish. Using the method of images, we can construct two SW configurations satisfying the boundary conditions in PM [Fig. 1(a),(c)] and SC [Fig. 1(b),(d)]. It can be seen that the distributions of dipoles and their images are different for **q** and $-$**q**, which correspond to different SW energies (or frequencies). For more detailed but greatly simplified explanations, see Appendix B. All of this makes the FM/PM(SC) system similar to the system with the Dzyaloshinskii-Moriya interaction (DMI) [17,18] that provides breaking the chiral symmetry for SW propagation in FM media with bulk (see, for example, [19]) and interfacial inversion asymmetry [20–34]. Moreover, the replacement of the substrate from PM to SC or from SC to PM corresponds to a change in the sign of the DMI constant. Despite the effects of SW nonreciprocity in FM/SC structures [35–36] and systems of coupled ferromagnets [34,37–46] were investigated, so far there is no a theory for SW propagation in FM/PM and FM/SC systems — with taking into account the finiteness and temperature sensitivity of the London penetration depth in SC and PM susceptibility.

For FM/PM system, we assume that (1) the Curie point ($T_c$) of its PM component is significantly lower than that in the FM one and that (2) temperature $T$ in the system is higher than $T_c$ but close to it. As for FM/SC system, it is assumed that the critical temperature for the normal metal-SC transition (again, denoted by $T_c$) in the SC component is higher than $T$. As we will show, the dipolar contribution to the SW nonreciprocity in the FM/PM system significantly exceeds that contribution due to the skin effect. So, one can neglect the electrical conductivity of the PM substrate. The interfacial exchange interaction that occurs between the film and the substrate can be suppressed by placing a thin dielectric layer between them. We also find that the effective DMI constant, which is of a dipole nature [42] in our case, is sensitive to changes in the *temperature-dependent* PM susceptibility (or the London penetration depth in SC) of the substrate. Thus, it becomes possible to control the SW frequency nonreciprocity by varying $T$ near $T_c$, which can be useful for applications. Currently, the possibility of controlling the DMI [47–63], spin waves [64–67] and SW nonreciprocity [42,68–70] is of considerable interest. It is also feasible to stabilize skyrmions [71] in such hybrid systems, using the same way as one does it in systems with strong enough interfacial DMI [72].

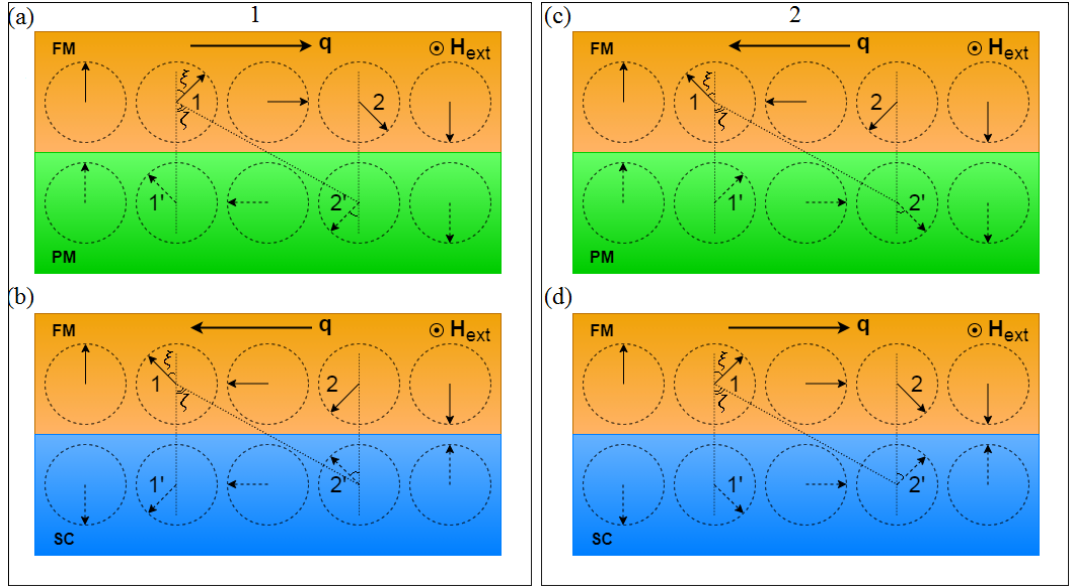

FIG. 1. Schematic for SW propagation along directions indicated by orientation of the SW wave vector **q** in FM/PM and FM/SC systems exposed by a magnetic field **H**$_{ext}$. The solid and dashed arrows indicate the SW chirality in the FM film and the SW image induced in PM (a), (c) and SC (b), (d) substrates, respectively. Here $\zeta$ is the angle between the normal to the film plane and the vector which connects the centers of dipoles 1 and 2', $\xi$ is the angle between the orientation of the dipole moment 1 and the normal to the film plane. The SW configurations shown in panel 1 only correspond to the minimal dipolar energy

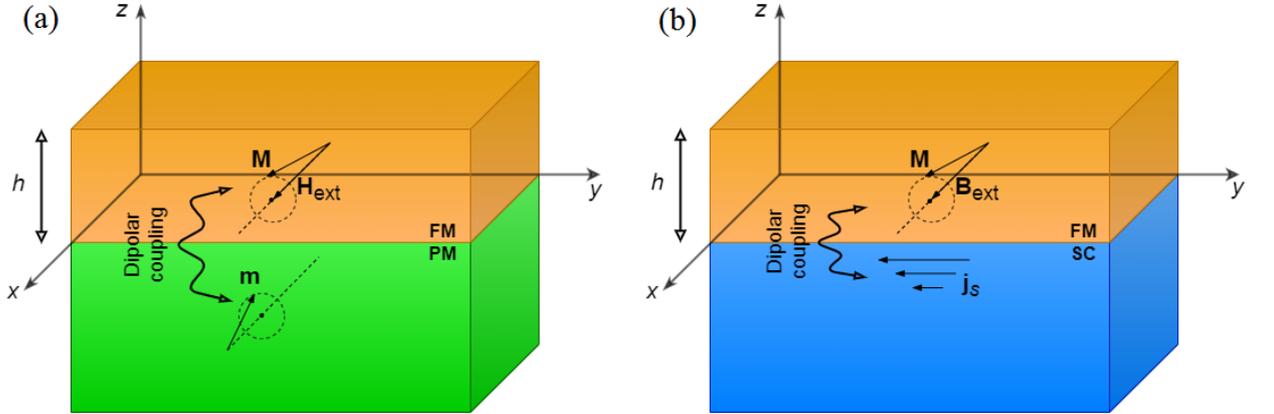

FIG. 2. Schematic representation of FM/PM (a) and FM/SC (b) systems. The FM film and PM (SC) substrate are coupled by interlayer dipolar interaction. The magnetization **M** in a spin wave that propagate in the FM film induces the magnetization **m** in PM substrate (a) or the superconducting current with density of **j**$_s$ in SC substrate (b), which affect back the SW propagation. **H**$_{ext}$ or **B**$_{ext}$ is the external magnetic field or magnetic flux density, respectively.

The article is organized as follows. In Section II, the exact geometry is presented for the systems we study and the starting equations have been derived. In Section III, we simulate magnetic fields generated by nonuniform magnetization in the FM. Section IV presents the calculation of the SW dispersion relation, from which the frequency shifts between counter propagating spin waves have been obtained in the systems under study. Section V presents illustrations for the results of our simulations. In Appendix A we treat the case in which the nonlocality of magnetization in the PM substrate cannot be neglected, and in Appendix B we present a qualitative explanation for the SW frequency nonreciprocity in the FM/PM and FM/SC systems.

## II. GEOMETRY OF THE SYSTEM AND STARTING EQUATIONS

As for the exact geometry of our system, we place the FM film at $0 < z < h$, while a PM or SC substrate is at $z < 0$ (Fig. 2) and $\mathbf{H}_{ext}$ or $\mathbf{B}_{ext}$, which is assumed to act on the FM only, is directed along the $x$ axis. In Sec. V, we will discuss the possibility of neglecting the action of $\mathbf{H}_{ext}$ or $\mathbf{B}_{ext}$ on the substrate. In our geometry, the normal vector $\mathbf{n}$ coincides with the unit vector of the $z$ axis. Nonuniform magnetization $\mathbf{M}$, in the FM film induces the magnetization $\mathbf{m}$ and superconducting current $\mathbf{j}_s$ in PM and SC substrates, respectively, which affect $\mathbf{M}$ back by their magnetic fields produced. We assume that the time needed to establish the equilibrium between the film and substrate is much shorter than $\omega^{-1}$, so that the magnetostatic approximation [2] could be used to calculate the magnetic fields generated.

### A. FM/PM

The total free energy of an FM/PM system can be divided into three contributions to it, which are (1) the free energy of the FM film ($F_0$), (2) the free energy of the PM substrate ($F_-$), and (3) the free energy of space at $z > h$ above the FM film ($F_+$):

$$F = F_0 + F_- + F_+, \qquad (2)$$

$$F_0 = \int_{-\infty}^{+\infty} \int_0^h \left( \frac{L_0^2}{2} \sum_{i=1}^{2} \left( \frac{\partial \mathbf{M}}{\partial x_i} \right)^2 - \frac{1}{2} K M_z^2 - [\mathbf{M} \cdot (\mathbf{H}_{ext} + \mathbf{H}_0)] - \frac{\mathbf{H}_0^2}{8\pi} \right) d\boldsymbol{\rho} dz, \qquad (3)$$

$$F_- = \int_{-\infty}^{+\infty} \int_{-\infty}^{0} \left( \frac{l_0^2}{2} \sum_{i=1}^{3} \left( \frac{\partial \mathbf{m}}{\partial x_i} \right)^2 + \frac{1}{2\chi} \mathbf{m}^2 - (\mathbf{m} \cdot \mathbf{H}_-) - \frac{\mathbf{H}_-^2}{8\pi} \right) d\boldsymbol{\rho} dz, \qquad (4)$$

$$F_+ = -\int_{-\infty}^{+\infty} \int_h^{+\infty} \frac{\mathbf{H}_+^2}{8\pi} d\boldsymbol{\rho} dz, \qquad (5)$$

where $K > 0$ is the dimensionless constant of magnetic anisotropy, $\boldsymbol{\rho} = (x, y)$ the two-dimensional coordinate vector, $\chi = C/(T - T_c)$ the substrate susceptibility and $C$ the Curie constant. The first terms in $F_0$ and $F_-$ describe the exchange interaction in the film and substrate, which is given by the constants $L_0$ and $l_0$, respectively. The relationship between $L_0$ and exchange stiffness $A$ is $L_0 = (2A/M_s^2)^{1/2}$, where $M_s$ is saturation magnetization (similarly for $l_0$). The quantities of $\mathbf{H}_-$ ($z < 0$), $\mathbf{H}_0$ ($0 < z < h$), and $\mathbf{H}_+$ ($z > h$) are the magnetic fields, induced by magnetizations of the film and substrate. We consider the film to be thin enough ($h \ll L_0$) [23], so that the dependence of the magnetization $\mathbf{M}$ on the coordinate $z$ normal to the film plane could be discarded. In what follows we will have revealed that the condition of $qh \ll 1$ should be also valid, in order to neglect that dependence.

As a magnetic field of any kind, the magnetic field produced by $\mathbf{M}$, i.e.,

$$\mathbf{H}(\boldsymbol{\rho}, z) = \begin{cases} \mathbf{H}_-(\boldsymbol{\rho}, z), z < 0 \\ \mathbf{H}_0(\boldsymbol{\rho}, z), 0 < z < h \\ \mathbf{H}_+(\boldsymbol{\rho}, z), z > h \end{cases} \quad (6)$$

can be described in terms of the scalar potential $\varphi(\boldsymbol{\rho}, z)$, so that $\mathbf{H}(\boldsymbol{\rho}, z) = -\nabla\varphi(\boldsymbol{\rho}, z)$. Then, the Maxwell's equations in the magnetostatic limit, that are, $\text{div}\mathbf{B} = 0$ and $\text{rot}\mathbf{H} = 0$, where $\mathbf{B} = \mathbf{H} + 4\pi\mathfrak{M}$, $\mathfrak{M}(\boldsymbol{\rho}, z) = \mathbf{m}(\boldsymbol{\rho}, z)\theta(-z) + \mathbf{M}(\boldsymbol{\rho})\theta(z)\theta(h - z)$, and $\theta(t)$ is Heaviside step function can be transformed into the Poisson's equation,

$$\Delta\varphi(\boldsymbol{\rho}, z) = 4\pi \, \text{div}\mathfrak{M}(\boldsymbol{\rho}, z), \quad (7)$$

which must be supplemented by the conditions that $\varphi$ and $B_z$ are continuous at $z = 0$ and $z = h$. A solution of Eq. (7), which satisfies the boundary conditions, can be written as

$$\mathbf{H}(\boldsymbol{\rho}, z) = \mathbf{H}^M(\boldsymbol{\rho}, z) + \mathbf{H}^m(\boldsymbol{\rho}, z), \quad (8)$$

$$\mathbf{H}^M(\boldsymbol{\rho}, z) = \int_{-\infty}^{+\infty}\int_0^h \text{div}'\mathbf{M}(\boldsymbol{\rho}')\nabla\frac{1}{|\mathbf{r}-\mathbf{r}'|}d\boldsymbol{\rho}'dz' - \int_{-\infty}^{+\infty} M_z(\boldsymbol{\rho}')\nabla\frac{1}{|\mathbf{r}-\mathbf{r}'|}\Big|_{z'=0}^{z'=h}d\boldsymbol{\rho}', \quad (9)$$

$$\mathbf{H}^m(\boldsymbol{\rho}, z) = \int_{-\infty}^{+\infty}\int_{-\infty}^0 \text{div}'\mathbf{m}(\boldsymbol{\rho}',z')\nabla\frac{1}{|\mathbf{r}-\mathbf{r}'|}d\boldsymbol{\rho}'dz' - \int_{-\infty}^{+\infty} m_z(\boldsymbol{\rho}',0)\nabla\frac{1}{|\mathbf{r}-\mathbf{r}'|}\Big|_{z'=0}d\boldsymbol{\rho}', \quad (10)$$

where $\nabla$ and $\text{div}'$ act on the $\mathbf{r}$ and $\mathbf{r}'$ coordinates, respectively. We see that both volume and surface magnetic charges in the film and substrate contribute to $\mathbf{H}(\boldsymbol{\rho}, z)$. From the condition that $\delta F/\delta\mathbf{m} = 0$, we obtain the following equation for $\mathbf{m}$ along with the boundary condition for that quantity:

$$\Delta\mathbf{m} - \frac{1}{l_m^2}\mathbf{m} = -\frac{\mathbf{H}_-}{l_0^2}, \qquad \frac{\partial\mathbf{m}}{\partial z}(z = 0) = 0, \quad (11)$$

where $l_m = l_0\sqrt{\chi}$. A solution of Eq. (11) can be represented as

$$\mathbf{m}(\boldsymbol{\rho}, z) = \mathbf{m}_0(\boldsymbol{\rho}, z) + \frac{1}{4\pi l_0^2}\int_{-\infty}^{+\infty}\int_{-\infty}^0 \frac{\exp(-|\mathbf{r}-\mathbf{r}'|/l_m)}{|\mathbf{r}-\mathbf{r}'|}\mathbf{H}_-(\mathbf{r}')d\boldsymbol{\rho}'dz', \quad (12)$$

where $\mathbf{m}_0(\boldsymbol{\rho}, z)$ is the homogeneous solution of Eq. (11). We note that taking into account exchange in the substrate gives a non-local dependence of $\mathbf{m}$ on $\mathbf{H}_-$. Using Eqs. (11) and the conditions that $\varphi$ and $B_z$ are continuous, we reduce the total free energy of the system to

$$F = \int_{-\infty}^{+\infty}\int_0^h \left(\frac{L_0^2}{2}\sum_{i=1}^2\left(\frac{\partial\mathbf{M}}{\partial x_i}\right)^2 - \frac{1}{2}KM_z^2 - \left[\mathbf{M}\cdot\left(\mathbf{H}_{\text{ext}} + \frac{\mathbf{H}_0}{2}\right)\right]\right)d\boldsymbol{\rho}dz. \quad (13)$$

We can see that the free energy of the substrate $F_-$ and the energy of the field $\mathbf{H}(\boldsymbol{\rho}, z)$ cancel each other. Under this circumstance, the effective field acting on the spin wave, which should be known to calculate the SW spectrum and thus the SW frequency nonreciprocity, can be reduced to

$$\mathbf{H}_{\text{eff}} = -\frac{\delta F}{\delta\mathbf{M}} = \mathbf{H}_{\text{ext}} + KM_z\mathbf{n} + L_0^2\Delta\mathbf{M} + \mathbf{H}_0. \quad (14)$$

## B.FM/SC

It is convenient to consider the free energy of the FM/SC system as a function of magnetic flux density, **B**. Except such a modification, equations for the free energy of the FM/SC system are identical to those for the FM/PM system given by Eqs. (3–5):

$$F_0 = \int_{-\infty}^{+\infty} \int_0^h \left( \frac{L_0^2}{2} \sum_{i=1}^2 \left(\frac{\partial \mathbf{M}}{\partial x_i}\right)^2 - \frac{1}{2} K M_z^2 - [\mathbf{M} \cdot (\mathbf{B}_{\text{ext}} + \mathbf{B}_0)] + \frac{\mathbf{B}_0^2}{8\pi} \right) d\boldsymbol{\rho} dz, \quad (15)$$

$$F_- = \frac{1}{8\pi} \int_{-\infty}^{+\infty} \int_{-\infty}^0 [\mathbf{B}_-^2 + l_s^2 (\text{rot}\mathbf{B}_-)^2] d\boldsymbol{\rho} dz, \quad (16)$$

$$F_+ = \int_{-\infty}^{+\infty} \int_h^{+\infty} \frac{\mathbf{B}_+^2}{8\pi} d\boldsymbol{\rho} dz, \quad (17)$$

where $\mathbf{B}_{\text{ext}}$ is the external magnetic field applied to the FM and $l_s$ the London penetration depth. The first and second terms in Eq. (16) represent the energies of the magnetic field and superconducting current, respectively. To describe the magnetic induction field induced by $\mathbf{M}(\boldsymbol{\rho})$, i.e.,

$$\mathbf{B}(\boldsymbol{\rho}, z) = \begin{cases} \mathbf{B}_-(\boldsymbol{\rho}, z), z < 0 \\ \mathbf{B}_0(\boldsymbol{\rho}, z), 0 < z < h \\ \mathbf{B}_+(\boldsymbol{\rho}, z), z > h \end{cases} \quad (18)$$

it is convenient to use the vector potential $\mathbf{A}(\boldsymbol{\rho}, z)$, so that $\mathbf{B}(\boldsymbol{\rho}, z) = \text{rot}\mathbf{A}(\boldsymbol{\rho}, z)$. Combining the Maxwell's equations, that are, $\text{rot}\mathbf{B}(\boldsymbol{\rho}, z) = 4\pi \, \text{rot}[\mathbf{M}(\boldsymbol{\rho})\theta(z)\theta(h-z)] + 4\pi\theta(-z)\mathbf{j}_s(\boldsymbol{\rho}, z)/c$ and $\text{div}\mathbf{B}(\boldsymbol{\rho}, z) = 0$, with the London one, that is $\mathbf{j}_s(\boldsymbol{\rho}, z) = -c\mathbf{A}_-(\boldsymbol{\rho}, z)/(4\pi l_s^2)$ for $z < 0$, we obtain for the vector potential that

$$\Delta \mathbf{A} = -4\pi \left( \theta(z)\theta(h-z) \, \text{rot}\mathbf{M} + [\delta(z) - \delta(h-z)](\mathbf{n} \times \mathbf{M}) - \frac{1}{4\pi l_s^2} \theta(-z)\mathbf{A} \right). \quad (19)$$

As usual, the tangential components of $\mathbf{A}(\boldsymbol{\rho}, z)$ and the normal component of $\mathbf{B}(\boldsymbol{\rho}, z)$ should be continuous at the interfaces $z = 0$ and $z = h$. In addition, the London equation requires the London gauge of $\mathbf{A}(\boldsymbol{\rho}, z)$, so that $\text{div}\mathbf{A} = 0$ and $[\mathbf{n} \cdot \mathbf{A}_-(z = 0)] = 0$. Eq. (19) can be transformed into the following integral equation:

$$\mathbf{A}(\boldsymbol{\rho}, z) = \mathbf{A}^*(\boldsymbol{\rho}, z) + \mathbf{A}^M(\boldsymbol{\rho}, z) + \mathbf{A}^s(\boldsymbol{\rho}, z), \quad (20)$$

$$\mathbf{A}^M(\boldsymbol{\rho}, z) = \int_{-\infty}^{+\infty} \int_0^h \frac{1}{|\mathbf{r} - \mathbf{r}'|} \text{rot}\mathbf{M}(\boldsymbol{\rho}') d\boldsymbol{\rho}' dz' - \int_{-\infty}^{+\infty} [\mathbf{n} \times \mathbf{M}(\boldsymbol{\rho}')] \frac{1}{|\mathbf{r} - \mathbf{r}'|} \Big|_{z'=0}^{z'=h} d\boldsymbol{\rho}', \quad (21)$$

$$\mathbf{A}^s(\boldsymbol{\rho}, z) = -\frac{1}{4\pi l_s^2} \int_{-\infty}^{+\infty} \int_{-\infty}^0 \frac{1}{|\mathbf{r} - \mathbf{r}'|} \mathbf{A}_-(\boldsymbol{\rho}', z') d\boldsymbol{\rho}' dz'. \quad (22)$$

The term $\mathbf{A}^*$ is a solution of Eq. (19) with zero right-hand side, whose introducing is necessary to satisfy the London gauge. It can be shown that $\text{rot}\mathbf{A}^* = 0$, so the potential $\mathbf{A}^*$ is not associated with the flow of currents and does not contribute to $\mathbf{B}(\boldsymbol{\rho}, z)$. From the above Maxwell's equations and the London one we can get the equation for $\mathbf{A}_-$:

$$\Delta \mathbf{A}_- - \frac{1}{l_s^2}\mathbf{A}_- = 0. \tag{23}$$

Using Eq. (23), as well as the conventional boundary conditions, we obtain expressions similar to Eqs. (13–14) for the free energy and for the effective field:

$$F = \int_{-\infty}^{+\infty}\int_0^h \left(\frac{L_0^2}{2}\sum_{i=1}^2\left(\frac{\partial \mathbf{M}}{\partial x_i}\right)^2 - \frac{1}{2}KM_z^2 - \left[\mathbf{M}\cdot\left(\mathbf{B}_{\text{ext}} + \frac{\mathbf{B}_0}{2}\right)\right]\right)d\boldsymbol{\rho}dz. \tag{24}$$

$$\mathbf{B}_{\text{eff}} = -\frac{\delta F}{\delta \mathbf{M}} = \mathbf{B}_{\text{ext}} + KM_z\mathbf{n} + L_0^2\Delta\mathbf{M} + \mathbf{B}_0. \tag{25}$$

Note here that, similar to that in FM/PM system, the magnetic energy and the energy of superconducting current cancel each other.

### III. CALCULATION OF THE GENERATED MAGNETIC FIELD

In this section, by using the Fourier transform method, we calculate the generated magnetic field $\mathbf{H}$ in an FM/PM system or its analog in a FM/SC system, $\mathbf{B}$. This quantity has to be calculated for determining the effective field that acts on a spin wave in the FM film. The Fourier transform for the SW magnetization $\mathbf{M}(\boldsymbol{\rho})$ is the plane wave expansion, i.e.,

$$\mathbf{M}(\boldsymbol{\rho}) = \sum_{\mathbf{q}} \mathbf{M}(\mathbf{q})e^{i(\mathbf{q}\cdot\boldsymbol{\rho})}, \tag{26}$$

where $\mathbf{M}(\mathbf{q})$ has the form

$$\mathbf{M}(\mathbf{q}) = \frac{1}{S}\int_{-\infty}^{+\infty}\mathbf{M}(\boldsymbol{\rho})e^{-i(\mathbf{q}\cdot\boldsymbol{\rho})}d\boldsymbol{\rho}. \tag{27}$$

Here $q_i = 2\pi N_i/L_i$ is $i$-th ($i = x, y$) component of the SW vector $\mathbf{q}$, $N_i \in \mathbb{Z}$, $S = L_xL_y$, $L_i$ is longitudinal system dimension.

### A. FM/PM

The Fourier transform of magnetization $\mathbf{m}(\boldsymbol{\rho}, z)$ in a PM substrate derived from Eq. (12) is

$$\mathbf{m}(\mathbf{q}, z) = S\int_{-\infty}^0 X(\mathbf{q}, z, z')\mathbf{H}_-(\mathbf{q}, z')dz', \tag{28}$$

where $X(\mathbf{q}, z, z')$ is the nonlocal substrate susceptibility density, which has the form

$$X(\mathbf{q}, z, z') = \frac{1}{2l_0^2 S\lambda_m}\left(e^{\lambda_m(z+z')} + e^{-\lambda_m|z-z'|}\right), \quad \lambda_m = \sqrt{q^2 + \frac{1}{l_m^2}}. \tag{29}$$

The field $\mathbf{H}_-(\mathbf{q}, z)$ in Eq. (28) can be found from the integral equation

$$H_{-i}(\mathbf{q},z) - \int_{-\infty}^{0} g_{ik}(\mathbf{q},z,z')H_{-k}(\mathbf{q},z')dz' = H_{-i}^{M}(\mathbf{q},z), \qquad (30)$$

which is obtainable from Eqs. (8–10,28). $\mathbf{H}_{-}^{M}(\mathbf{q},z)$ on the right side of Eq. (30) denotes the field (9) in the area $z < 0$ and has the form

$$\mathbf{H}_{-}^{M}(\mathbf{q},z) = 2\pi(1 - e^{-qh})e^{qz}\left(\frac{i[\mathbf{q}\cdot\mathbf{M}(\mathbf{q})]}{q} + M_z(\mathbf{q})\right)\left(\frac{i\mathbf{q}}{q} + \mathbf{n}\right). \qquad (31)$$

In Eq. (30), the components of the $\hat{g}(\mathbf{q},z,z')$ tensor are given in Appendix A. In these considerations we assume that $q l_0/2\sqrt{\pi} \ll 1$ and $q l_0 \sqrt{\chi} \ll 1$, so that $\mathbf{m} = \chi \mathbf{H}_{-}$ and $\mathrm{div}\,\mathbf{m} \to 0$. In this local limit we get that

$$g_{ik}(\mathbf{q},z,z') = \begin{cases} 0; & i,k = x,y \\ -\dfrac{4\pi\chi i q_i}{q} e^{qz}\delta(z'); & i = x,y; k = z \\ 0; & i = z; k = x,y \\ -4\pi\chi e^{qz}\delta(z'); & i,k = z \end{cases}. \qquad (32)$$

Then Eq. (30) becomes linear and its solution reads

$$\mathbf{H}_{-}(\mathbf{q},z) = (1 - \kappa)\mathbf{H}_{-}^{M}(\mathbf{q},z), \qquad (33)$$

where $\kappa = 2\pi\chi/(1 + 2\pi\chi)$. Thus, if the substrate is an ideal PM ($\chi \to \infty$), the magnetic field in its region is completely suppressed by the induced surface charges, while the magnetization remains finite, $\mathbf{m}(\mathbf{q},z) = \mathbf{H}_{-}^{M}(\mathbf{q},z)/2\pi$. With taking into account $\mathbf{H}_{-}(\mathbf{q},z)$ found, one gets the rest of the fields from Eqs. (8–10):

$$\mathbf{H}_{+}(\mathbf{q},z) = \mathbf{H}_{+}^{M}(\mathbf{q},z) + \mathbf{H}_{z>0}^{m}(\mathbf{q},z), \qquad (34)$$

$$\mathbf{H}_{0}(\mathbf{q},z) = \mathbf{H}_{0}^{M}(\mathbf{q},z) + \mathbf{H}_{z>0}^{m}(\mathbf{q},z), \qquad (35)$$

where the contributions from the film charges are as follows

$$\mathbf{H}_{+}^{M}(\mathbf{q},z) = 2\pi(e^{qh} - 1)e^{-qz}\left(\frac{i[\mathbf{q}\cdot\mathbf{M}(\mathbf{q})]}{q} - M_z(\mathbf{q})\right)\left(\frac{i\mathbf{q}}{q} - \mathbf{n}\right), \qquad (36)$$

$$\mathbf{H}_{0}^{M}(\mathbf{q},z) = \frac{2\pi i \mathbf{q}}{q}\left(\frac{i[\mathbf{q}\cdot\mathbf{M}(\mathbf{q})]}{q}(2 - e^{qz}e^{-qh} - e^{-qz}) - M_z(\mathbf{q})(e^{qz}e^{-qh} - e^{-qz})\right)$$

$$+ 2\pi\mathbf{n}\left(-\frac{i[\mathbf{q}\cdot\mathbf{M}(\mathbf{q})]}{q}(e^{qz}e^{-qh} - e^{-qz}) - M_z(\mathbf{q})(e^{qz}e^{-qh} + e^{-qz})\right), \qquad (37)$$

while the contributions from the substrate charges are $\mathbf{H}_{+}^{m}(\mathbf{q},z) = \mathbf{H}_{0}^{m}(q,z) \equiv \mathbf{H}_{z>0}^{m}(\mathbf{q},z)$, where

$$\mathbf{H}_{z>0}^{m}(\mathbf{q},z) = -4\pi J_m e^{-qz}\left(\frac{i[\mathbf{q}\cdot\mathbf{M}(\mathbf{q})]}{q} + M_z(\mathbf{q})\right)\left(\frac{i\mathbf{q}}{q} - \mathbf{n}\right) \qquad (38)$$

and

$$J_m = \frac{\kappa}{2}(1 - e^{-qh}). \quad (39)$$

We see that the field $\mathbf{H}_0$ depends on $z$ and thus $\mathbf{M}$ depends on $z$ as well, despite our initial requirement. To avoid this problem, the field $\mathbf{H}_0(\mathbf{q}, z)$ has to be averaged over the FM thickness:

$$\mathbf{H}_0(\mathbf{q}) \equiv \frac{1}{h}\int_0^h \mathbf{H}_0(\mathbf{q}, z)dz = \mathbf{H}_0^M(\mathbf{q}) + \mathbf{H}_{z>0}^m(\mathbf{q}), \quad (40)$$

$$\mathbf{H}_0^M(\mathbf{q}) = -\frac{4\pi}{qh}\left(\frac{[\mathbf{q}\cdot\mathbf{M}(\mathbf{q})]}{q^2}(qh - 1 + e^{-qh})\mathbf{q} + M_z(\mathbf{q})(1 - e^{-qh})\mathbf{n}\right), \quad (41)$$

$$\mathbf{H}_{z>0}^m(\mathbf{q}) = -\frac{4\pi J_m}{qh}(1 - e^{-qh})\left(\frac{i[\mathbf{q}\cdot\mathbf{M}(\mathbf{q})]}{q} + M_z(\mathbf{q})\right)\left(\frac{i\mathbf{q}}{q} - \mathbf{n}\right). \quad (42)$$

The condition that $qh \ll 1$, along with $h \ll L_0$, makes it possible to neglect the dependence of $\mathbf{M}$ on $z$. As expected, $\mathbf{H}_0^M(\mathbf{q})$ after expansion in $qh$ reduces to the demagnetizing field, $-4\pi M_z(\mathbf{q})\mathbf{n}$. Note also that Eqs. (33–38) can be obtained using the method of images even if the substrate is not an ideal PM [15].

### A. FM/SC

According to Eq. (23), the Fourier transform of $\mathbf{A}_-$ is $\mathbf{A}_-(\mathbf{q}, z) = \mathbf{A}_-(\mathbf{q}, 0)e^{\lambda_s z}$, where $\lambda_s = (q^2 + 1/l_s^2)^{1/2}$. Moreover, according to the London gauge, there should be $[\mathbf{q}\cdot\mathbf{A}_-(\mathbf{q}, z)] = [\mathbf{n}\cdot\mathbf{A}_-(\mathbf{q}, z)] = 0$. Therefore, from Eqs. (20–22), one gets $\mathbf{A}_-(\mathbf{q}, z)$ as well as this quantity in other regions of space, which enables us to retrieve the Fourier transform of the field $\mathbf{B}(\mathbf{q}, z)$ in different regions of the system:

$$\mathbf{B}_-(\mathbf{q}, z) = 2\pi(1 - e^{-qh})\left(1 - \frac{\lambda_s - q}{\lambda_s + q}\right)e^{\lambda_s z}\left(\frac{i[\mathbf{q}\cdot\mathbf{M}(\mathbf{q})]}{q} + M_z(\mathbf{q})\right)\left(\frac{i\lambda_s \mathbf{q}}{q^2} + \mathbf{n}\right), \quad (43)$$

$$\mathbf{B}_+(\mathbf{q}, z) = \mathbf{B}_+^M(\mathbf{q}, z) + \mathbf{B}_{z>0}^s(\mathbf{q}, z), \quad (44)$$

$$\mathbf{B}_0(\mathbf{q}, z) = \mathbf{B}_0^M(\mathbf{q}, z) + \mathbf{B}_{z>0}^s(\mathbf{q}, z), \quad (45)$$

$$\mathbf{B}_{z>0}^s(\mathbf{q}, z) = -4\pi J_s e^{-qz}\left(\frac{i[\mathbf{q}\cdot\mathbf{M}(\mathbf{q})]}{q} + M_z(\mathbf{q})\right)\left(\frac{i\mathbf{q}}{q} - \mathbf{n}\right). \quad (46)$$

Note that, of course, $\mathbf{B}_+^M(\mathbf{q}, z) = \mathbf{H}_+^M(\mathbf{q}, z)$ [see Eq. (36)] and $\mathbf{B}_0^M(\mathbf{q}, z) = \mathbf{H}_0^M(\mathbf{q}, z) + 4\pi\mathbf{M}(\mathbf{q})$ [see Eq. (37)], and so Eq. (46) is equivalent to Eq. (38), except replacing $J_m$ by $J_s$, where

$$J_s = -\frac{1}{2}\cdot\frac{\lambda_s - q}{\lambda_s + q}(1 - e^{-qh}). \quad (47)$$

In the case of an ideal SC ($l_s \to 0$) there will be $\mathbf{B}_-(\mathbf{q}, z) \to 0$ and $[\mathbf{n}\cdot\mathbf{B}_0(\mathbf{q}, 0)] \to 0$, which can also be obtained by using the method of images.

## IV. DYNAMIC PROBLEM: SW DISPERSION CURVES

As usual, in order to clarify the magnetization dynamics in the systems under study with calculation of the SW dispersion curves, we use the linearized Landau–Lifshitz–Gilbert (LLG) equation that contains the effective field given by Eqs. (14) an (24). Under the condition that $H_{ext}^0 > K - 4\pi$, where $H_{ext}^0$ is the absolute value of $\mathbf{H}_{ext}$ or $\mathbf{B}_{ext}$, the SW dispersion relation found is $\omega(\mathbf{q}) = \omega'(\mathbf{q}) + i\omega''(\mathbf{q})$, where $\omega'(\mathbf{q})$ and $\omega''(\mathbf{q})$ are real and imaginary parts of the spectrum, respectively:

$$\omega'(\mathbf{q}) = -\frac{\Delta\omega'(\mathbf{q})}{2} + \gamma M_s \sqrt{\left(\frac{H_{ext}^0}{M_s} + L_0^2 q^2 + I_1\right)\left(\frac{H_{ext}^0}{M_s} + L_0^2 q^2 + I_2 - K\right)}, \quad (48)$$

$$\omega''(\mathbf{q}) = -\frac{\alpha\gamma\omega'(\mathbf{q})[2H_{ext}^0 + M_s(2L_0^2 q^2 - K + I_1 + I_2)]}{2\omega'(\mathbf{q}) + \Delta\omega'(\mathbf{q})}. \quad (49)$$

Where $\gamma > 0$ is the gyromagnetic ratio, $\alpha$ the damping parameter and

$$\Delta\omega'(\mathbf{q}) = \omega'(-\mathbf{q}) - \omega'(\mathbf{q}) = \frac{8\pi J \gamma M_s}{q^2 h}(1 - e^{-qh})\left(\mathbf{q} \cdot \left(\mathbf{n} \times \frac{\mathbf{M}_s}{M_s}\right)\right), \quad (50)$$

$$I_1 = \frac{4\pi q_y^2}{q^3 h}\left(qh - 1 + e^{-qh} - J(1 - e^{-qh})\right), \quad (51)$$

$$I_2 = \frac{4\pi}{qh}(1 - e^{-qh})(1 - J). \quad (52)$$

where the constant $J$ is equal to $J_m$ and $J_s$ in FM/PM and FM/SC systems, respectively. We see that the SW spectrum depends on SW propagation direction along the $y$-axis. This nonreciprocity is determined by the substrate PM susceptibility (or the London penetration depthin SC) and disappears in the limit $\chi \to 0$ ($l_s \to \infty$). The presence of nonreciprocity in the imaginary part of the spectrum indicates different life times for the counter propagating spin waves. For $q = 0$ from Eq. (48) and Eq. (49), as expected, we obtain the formulas corresponding to the uniform precession of magnetization in a longitudinally magnetized FM film [2], since $I_1 \to 0$, $I_2 \to 4\pi$ and $\Delta\omega' \to 0$. Also, for $\chi \to 0$ ($l_s \to \infty$), $L_0 = 0$, $K = 0$, $q_x = 0$ and $qh \ll 1$ we get the Damon-Eshbach mode, i.e. $\omega' = \gamma[H_{ext}^0(H_{ext}^0 + 4\pi H_{ext}^0 M_s + 8\pi^2 M_s^2 |q_y| h)]^{1/2}$ [8].

The applicability of the above magnetostatic approach is provided by the small penetration depth of the $\mathbf{H}_-$ compared to the skin depth in the PM. The skin depth $\delta$ of highly conductive metals (copper, aluminum) for $\omega \sim 10^{10} s^{-1}$ is of the order of $10^{-4}$ cm. Then the condition $q\delta \gg 1$ provides the smallness of the penetration depth of the $\mathbf{H}_-$ and is satisfied for $q \geq 10^5$ cm$^{-1}$. If $h \sim 10^{-7}$ cm (note that $h \ll L_0$ and $L_0 \sim 10^{-7} \div 10^{-6}$ cm), then the condition $q\delta \gg 1$ is compatible with $qh \ll 1$ for $q \sim 10^5 \div 10^6$ cm$^{-1}$. Thus, there is a range of SW wavelengths, within of which the magnetostatic approach is applicable.

# V. RESULTS AND DISCUSSIONS

Figures 3 and 4 show dispersion curves calculated according to Eq. (48) and Eq. (49) at $h = 3$ nm, $M_s = 800$ erg $G^{-1}$cm$^{-3}$, $L_0 = 20$ nm, $H_{\text{ext}}^0 = 800$ Oe, $K = 0$, $q_x = 0$, $\alpha = 0.01$ (FM = Ni$_{80}$Fe$_{20}$). The dashed orange lines in Fig. 3 correspond to the local limit (see Sec. III.A), and the solid orange and blue ones correspond to the exact solution (see Appendix A). We can see that for $\chi = 0$ ($l_s \to \infty$) the spin waves are reciprocal, i.e., $\omega(\mathbf{q}) = \omega(-\mathbf{q})$. In this case, $\omega$ does not depend on $l_0$ in the FM/PM system, and the exact solution coincides with the local limit. When $\chi \neq 0$ ($l_s \neq \infty$), the spectrum becomes nonreciprocal, and the SW frequencies depend on the direction of SW propagation along the $y$ axis. We see that $l_0 \neq 0$ gives a decrease of nonreciprocity, $|\Delta\omega'|$, which can be explained by lowering the nonlocal susceptibility of the substrate [see Eq. (29)]. The nonreciprocity disappears completely in the $l_0 \to \infty$ limit. As can be also seen from Fig. 3, the difference between the exact and approximate solution increases with $q_y$.

Comparing our results with the frequency shift due to DMI [23], we obtain the effective DMI constant [42]

$$D_{\text{eff}} = -\frac{2\pi J}{q^2 h}(1 - e^{-qh})M_s^2. \quad (53)$$

Then, for $q_x = 0$ the frequency shift $\Delta f(\mathbf{q}) = \Delta\omega'(\mathbf{q})/2\pi$ due to dipolar coupling is

$$\Delta f(\mathbf{q}) = -\frac{2\gamma D_{\text{eff}} q_y}{\pi M_s}. \quad (54)$$

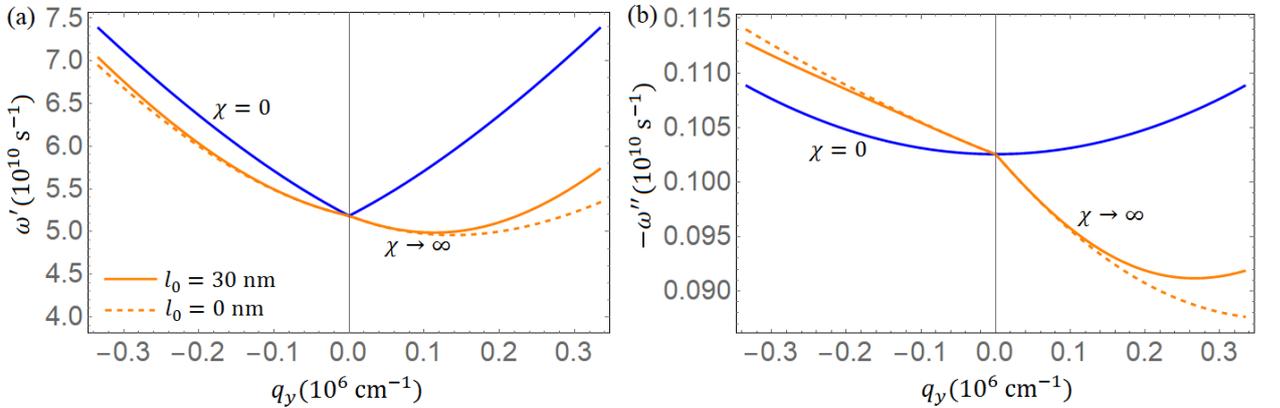

FIG. 3. Dispersion curves (a) $\omega'(\mathbf{q})$ and (b) $\omega''(\mathbf{q})$ for the FM/PM system, FM = Ni$_{80}$Fe$_{20}$. The solid blue curves correspond to $\chi = 0$. The orange curves correspond to $T = T_c$ ($\chi \to \infty$) at $l_0 = 0$ (dashed lines) and $l_0 = 30$ nm (solid lines)

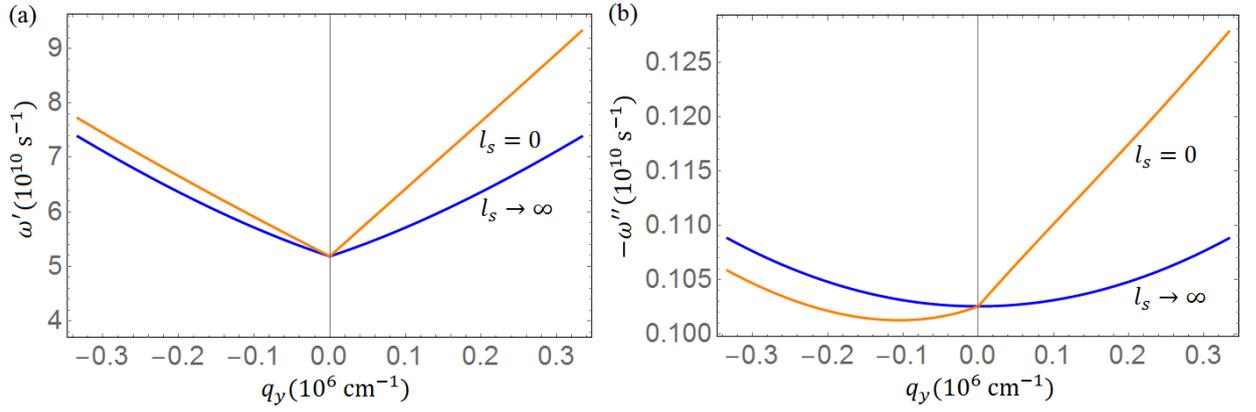

FIG. 4. Dispersion curves (a) $\omega'(\mathbf{q})$ and (b) $\omega''(\mathbf{q})$ for the FM/SC system, FM = $Ni_{80}Fe_{20}$. The blue curves correspond to $T = T_c$ ($l_s \to \infty$). The orange curves are for ideal SC ($l_s = 0$)

The quantity given by Eq. (54) can be measured experimentally by Brillouin spectroscopy [73]. Figures 5 and 6 show the dependences of $\Delta f(\mathbf{q})$ and $D_{\text{eff}}$ on $q_y$ for different types of substrates and temperatures (FM = $Ni_{80}Fe_{20}$, PM = Gd, SC = Pb). We used the following temperature dependence of the London penetration depth: $l_s(T) = l_s(0)(1 - T/T_c)^{-1/2}$. It can be seen that $\Delta f$ and $D_{\text{eff}}$ increase with decreasing temperature, which corresponds to the increase in the susceptibility and to the decrease in London penetration depth. A noticeable difference in $\Delta f$ for the two types of substrates can be explained by rather a large value of $l_s(0) = 39$ nm for Pb. If $4\pi\chi \gg 1$, $ql_0/2\sqrt{\pi} \ll 1$, $ql_0\sqrt{\chi} \ll 1$, $qh \ll 1$ (FM/PM system) and if $ql_s \ll 1$, $qh \ll 1$ (FM/SC system) then the quantities $D_{\text{eff}}$ and $\Delta f$ reach their maxima to be

$$|D_{\text{eff}}| \approx \pi M_s^2 h. \tag{55}$$

$$|\Delta f(\mathbf{q})| \approx 2\gamma M_s h |q_y|. \tag{56}$$

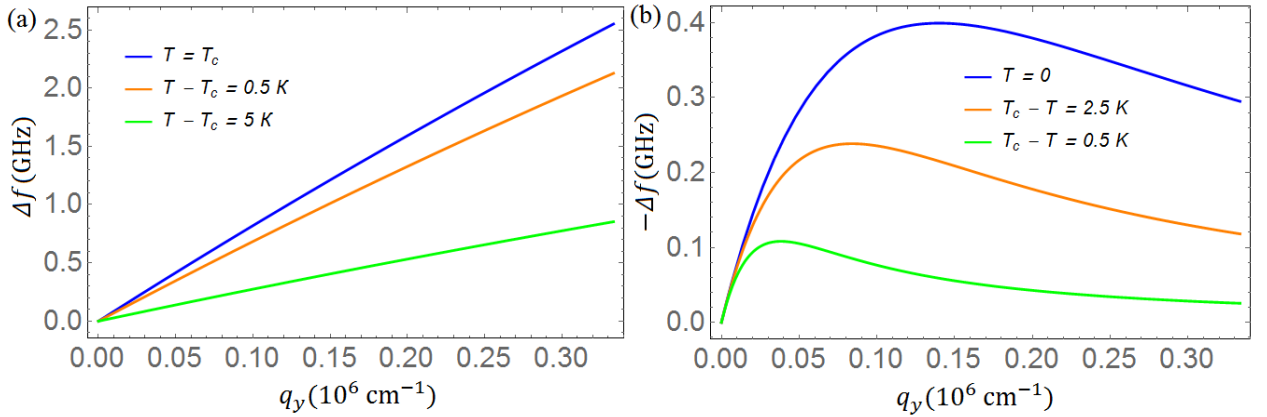

FIG. 5. Frequency shift vs. $q_y$ for (a) the FM/PM ($Ni_{80}Fe_{20}$/Gd) and (b) FM/SC ($Ni_{80}Fe_{20}$/Pb) systems. The following parameters were used in these calculations: $h = 3$ nm, $M_s = 800$ erg $G^{-1}cm^{-3}$, $q_x = 0$, $l_0 = 5$ nm, $l_s(0) = 39$ nm, $T_c$ for Gd and Pb is 293 K and 7.26 K, respectively

For a $Ni_{80}Fe_{20}$ film, $D_{\text{eff}}$ in these approximations reaches the value of 0.6 erg/cm$^2$. To increase the effective DMI, as can be seen, it is necessary to use materials with a high saturation magnetization $M_s$ and with film thickness as large as possible. Note that the frequency shift given by Eq. (56) is equal to that quantity in an FM/nonmagnetic metal system and can be associated with the skin effect [9–11]. However, this coincidence takes place only in the limit of infinite conductivity of the metal when the skin-depth thickness is equal to zero. For metals with finite conductivity, as we have shown earlier, the skin effect in the problem under consideration can be neglected, and so, the contribution of interlayer dipolar coupling is dominating in the nonreciprocity.

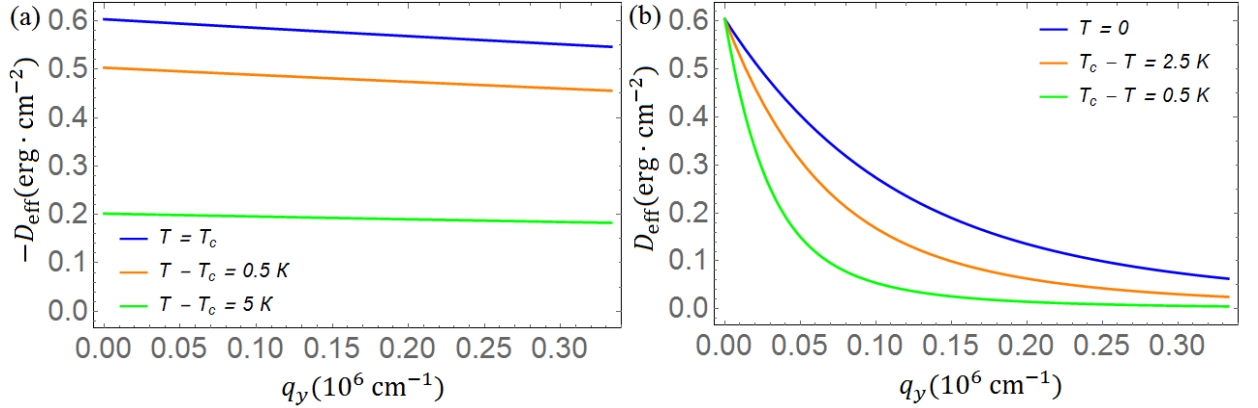

FIG. 6. Effective DMI constant vs. $q_y$ for (a) the FM/PM ($Ni_{80}Fe_{20}$/Gd) and (b) FM/SC ($Ni_{80}Fe_{20}$/Pb) systems. The following parameters were used in these calculations: $h = 3$ nm, $M_s = 800$ erg G$^{-1}$cm$^{-3}$, $q_x = 0$, $l_0 = 5$ nm, $l_s(0) = 39$ nm, $T_c$ for Gd and Pb is 293 K and 7.26 K, respectively

Previously (see Sec. II), we neglected the action of the $\mathbf{H}_{ext}$ ($\mathbf{B}_{ext}$) on the substrate. Let us show that the action of this field on the substrate does not affect the SW nonreciprocity. The $\mathbf{H}_{ext}$ ($\mathbf{B}_{ext}$) will cause an additional contribution to the PM magnetization and to the superconducting current in the SC. In our model, this contribution to the PM magnetization will be uniform and not lead to the creation of additional dipole fields. In the case of the SC substrate, additional superconducting currents caused by an external field will create an approximately uniform additional dipole field in the region of ferromagnetic film. Then in the Eqs. (48), (49) it is necessary to add this additional field to $H_{ext}^0$. But this will not affect the SW nonreciprocity.

Thus, in the present work, we have determined the SW spectra in FM/PM and FM/SC systems. The spectra are nonreciprocal due to the dipolar interaction between the FM film and PM or SC substrates. The temperature sensitivity of the substrate susceptibility (London penetration depth) allows for controlling the frequency shift by varying $T$ near $T_c$, which can be useful for applications.

## ACKNOWLEDGMENTS

We acknowledge very useful discussions with N.I. Polushkin. The work was supported by the State Contract No. 0030-2021-0021 and Russian Foundation for Basic Research (Grant No. 20-02-00356). One of us (M.A. K.) thanks to the Foundation for the Advancement of Theoretical Physics and Mathematics "BASIS".

## APPENDIX A: THE CASE OF NONLOCAL PM

Here we will consider a situation in which one cannot neglect the nonlocal dependence of $\mathbf{m}$ on $\mathbf{H}_-$ in the PM substrate. Let us return to the analysis of Eq. (30). In general, the components of the tensor $\hat{g}(\mathbf{q}, z, z')$ have the form

$$g_{ik}(\mathbf{q}, z, z') = \begin{cases} \frac{4\pi q_i q_k}{\lambda_m} e^{\lambda_m z'} \left( \frac{\chi}{1+4\pi\chi} e^{\Lambda z} - \frac{1}{2q l_0^2 (\Lambda - q)} e^{qz} \right); i, k = x, y \\ -\frac{2\pi i q_i}{q \lambda_m l_0^2} e^{\lambda_m z'} e^{qz}; i = x, y; k = z \\ -\frac{4\pi i q_k}{\lambda_m} e^{\lambda_m z'} \left( \frac{\chi}{1+4\pi\chi} \Lambda e^{\Lambda z} - \frac{1}{2 l_0^2 (\Lambda - q)} e^{qz} \right); i = z; k = x, y \\ -\frac{2\pi}{\lambda_m l_0^2} e^{\lambda_m z'} e^{qz}; i, k = z \end{cases}, \quad (A1)$$

where $\Lambda = (q^2 + 1/l_m^2 + 4\pi/l_0^2)^{1/2}$. It is possible to seek for a solution to Eq. (30) by expanding $\mathbf{H}_-$ and $\mathbf{M}$ into a series in $\chi$. A similar approach was used earlier in [13], where the response of the PM substrate was found in the linear approximation. However, Eq. (30) also admits an exact solution. Since the $z$ coordinate in Eqs. (30–31, A1) is contained only in the exponential functions $e^{\Lambda z}$ and $e^{qz}$, we will seek the exact solution in the form

$$\mathbf{H}_-(\mathbf{q}, z) = \left( \frac{i[\mathbf{q} \cdot \mathbf{M}(\mathbf{q})]}{q} + M_z(\mathbf{q}) \right) \left( (\mathbf{c}_1 + d_1 \mathbf{n}) e^{qz} + (\mathbf{c}_2 + d_2 \mathbf{n}) e^{\Lambda z} \right), \quad (A2)$$

where $d_1$ and $d_2$ are constants depending only on $\mathbf{q}$; $\mathbf{c}_1$ and $\mathbf{c}_2$ are two-dimensional vectors (have only $x$- and $y$-components), also depending only on $\mathbf{q}$. After substituting Eq. (A2) into Eq.(30), we obtain

$$\mathbf{H}_-(\mathbf{q}, z) = \left( \frac{i[\mathbf{q} \cdot \mathbf{M}(\mathbf{q})]}{q} + M_z(\mathbf{q}) \right) \left( d_1 e^{qz} \left( \frac{i\mathbf{q}}{q} + \mathbf{n} \right) + d_2 e^{\Lambda z} \left( \frac{i\mathbf{q}}{\Lambda} + \mathbf{n} \right) \right), \quad (A3)$$

where the constants $d_1$ and $d_2$ have the form

$$d_1 = \frac{2\pi b_1 (1 - e^{-qh})}{a_2 b_1 + a_1 b_2}, \qquad d_2 = \frac{2\pi a_1 (1 - e^{-qh})}{a_2 b_1 + a_1 b_2}, \quad (A4)$$

$$a_1 = \frac{4\pi \chi q \Lambda}{(1 + 4\pi\chi)(\lambda_m + q)\lambda_m}, \qquad a_2 = 1 + \frac{2\pi \Lambda}{\lambda_m l_0^2 (\lambda_m + q)(\Lambda - q)}, \quad (A5)$$

$$b_1 = 1 - \frac{4\pi \chi q^2}{\lambda_m (1 + 4\pi\chi)(\lambda_m + \Lambda)}, \qquad b_2 = \frac{2\pi}{\lambda_m l_0^2 (\lambda_m + \Lambda)} \left( 1 + \frac{q^2}{\Lambda(\Lambda - q)} \right). \quad (A6)$$

From the equation for $\mathbf{H}_-$, we can find $\mathbf{H}_+$ and $\mathbf{H}_0$. The equations for these quantities coincide with Eq. (34) and Eq. (35), but now the constant $J_m$ takes the form

$$J_m = \frac{1}{2\lambda_m l_0^2} \left( \frac{d_1}{q + \lambda_m} \left( 1 + \frac{q}{q + \Lambda} \right) + \frac{d_2}{\lambda_m + \Lambda} \left( 1 + \frac{q^2}{\Lambda} \frac{1}{q + \Lambda} \right) \right). \quad (A7)$$

We also present the expression for the magnetization $\mathbf{m}(\mathbf{q}, z)$, obtained from Eq. (28):

$$\mathbf{m}(\mathbf{q},z) = \left(\frac{i[\mathbf{q}\cdot\mathbf{M}(\mathbf{q})]}{q} + M_z(\mathbf{q})\right)\left(\chi d_1 e^{qz}\left(\frac{i\mathbf{q}}{q} + \mathbf{n}\right) - \frac{d_2}{4\pi}e^{\Lambda z}\left(\frac{i\mathbf{q}}{\Lambda} + \mathbf{n}\right)\right.$$
$$\left. + \frac{q}{\lambda_m}e^{\lambda_m z}\left(\frac{i\mathbf{q}}{q}\left(\frac{d_2}{4\pi} - \chi d_1\right) + \mathbf{n}\left(\frac{d_2\Lambda}{4\pi q} - \chi d_1\right)\right)\right). \quad \text{(A8)}$$

Now the field $\mathbf{H}_-$ consists of two terms with different transverse scales: $1/q$ and $1/\Lambda$. The second scale appears only in the non-local limit. Its appearance is associated with volume charges in the PM, which arise at an arbitrarily small $l_0$ (div$\mathbf{m} \neq 0$). The term with scale $1/\Lambda$ in Eq. (A3) disappears in the local limit ($l_0 \to 0$). The "local" term (with a scale of $1/q$) refers to the corresponding magnetization term (A8) as $1/\chi$. Since $l_0 \sim 10^{-7} \div 10^{-6}$cm, then $\Lambda\delta \gg 1$. In the nonlocal limit, this provides the smallness of the $\mathbf{H}_-$ penetration depth into the substrate compared to the skin depth. All previously obtained results for the local case remain to bevalid, but with the replacement of Eq. (39) by Eq. (A7) for the constant $J_m$. For $ql_0/2\sqrt{\pi} \ll 1$ and $ql_0\sqrt{\chi} \ll 1$, we get the local limit discussed in Sec. III.

Note the importance of taking into account the nonlocality in the substrate. When $T$ is close enough to the $T_c$, the condition of $ql_0\sqrt{\chi} \ll 1$ is not satisfied any longer, and so, the nonlocality is essential. Moreover, the factor of $ql_0$ can be of the order of unity.

## APPENDIX B: QUALITATIVE EXPLANATION FOR THE SW FREQUENCY NONRECIPROCITY IN THE FM/PM AND FM/SC SYSTEMS

To explain the mechanism behind the SW frequency nonreciprocity in the FM/PM(SC) system, we present a simplified illustration of SW propagationin an FM film on PM and SC substrates. As mentioned earlier, the rigorous treatments can be found in Refs. [13,16]. It is assumed that the PM and SC are ideal, that is, the tangential (for PM) or normal (for SC) component of a magnetic field should vanish at the boundary between the FM and PM (SC). Figure 1 shows snapshots of spin waves that propagate along opposite directions in FM/PM (a, c) and FM/SC (b, d). Such spin waves, which we consider as magnetic dipoles whose orientation is variable in space (shown by the solid arrows), induce respectively different image dipole configurations in PM (SC) (shown by the dashed arrows).

Let us compare the dipole energies in a spin wave propagating in FM/PM [Fig.1(a),(c)] and FM/SC [Fig.1(b),(d)] for the cases $+\mathbf{q}$ and $-\mathbf{q}$ by considering the pairwise interaction energies between dipoles in FM and their images in PM (SC). As seen from the system symmetry, the interaction energies of dipoles 1, 2, … and their own images, 1', 2', … as well as of dipoles inside the film or their images inside the substrate (for example, 1 and 2, 1' and 2') are equal each to other in the cases $+\mathbf{q}$ and $-\mathbf{q}$. We see, however, that the interaction energies of dipoles in FM and images in PM (SC) from other dipoles in FM (for example, 1 and 2', 2 and 1') are different for $+\mathbf{q}$ and $-\mathbf{q}$ directions. Indeed, this energy for dipoles 1 and 2' (or 2 and 1') is the form

$$E_\pm = \frac{2\mu^2\sigma}{d^3}[3\cos^2(\zeta \pm \xi) - 1], \quad \text{(B1)}$$

where µ is the dipole moment, the same for all dipoles, $d$ is the distance between dipoles 1 and 2' or (2 and 1'), $\sigma = 1$ for FM/PM and $\sigma = -1$ for FM/SC, $\zeta$ is the angle between the normal to the film plane and the vector which connects the dipoles, and $\xi$ is the angle between the moment orientation and the normal to the film plane. In Eq. (B1), the indices "+" and "−" correspond to $+\mathbf{q}$ and $-\mathbf{q}$ directions. Then the difference $E_+ - E_-$ is proportional to $-\sigma \sin(2\xi)\sin(2\zeta)$. Since $0 < \zeta < \pi/2$ and $0 < \xi < \pi/2$, we have that $E_+ < E_-$ for FM/PM and $E_+ > E_-$ for FM/SC. Similar results can be obtained by comparing the energies of the remaining dipoles and images. Thus, panel 1 in Fig. 1 corresponds to preferable SW propagation and these directions are different for FM/PM and FM/SC.